\documentclass{article}
\usepackage{microtype}
\usepackage{graphicx}
\usepackage{subfigure}
\usepackage{booktabs}
\usepackage{hyperref}
\usepackage{breakurl}

\usepackage[accepted]{icml2023}

\usepackage{amsmath}
\usepackage{amssymb}
\usepackage{mathtools}
\usepackage{amsthm}

\usepackage[capitalize,noabbrev]{cleveref}

\theoremstyle{plain}

\theoremstyle{definition}

\theoremstyle{remark}

\usepackage[textsize=tiny]{todonotes}

\usepackage{xspace}

\newcommand*{\boxmin}{-\xspace}
\newcommand*{\boxmax}{+\xspace}

\def\<{\langle}
\def\>{\rangle}

\def\defeq{\coloneqq} 

\renewcommand*{\Box}{\operatorname{Box}}

\newcommand\GBox{\operatorname{Box}_G}

\newcommand{\I}{\mathcal{I}}
\newcommand{\hI}{\hat{\mathcal{I}}}
\newcommand{\hY}{\hat{Y}} 
\newcommand{\RR}{\mathbb{R}}

\newcommand{\hiddencomment}[1]{}

\icmltitlerunning{Answering Compositional Queries with Set-Theoretic Embeddings}

\begin{document}

\twocolumn[
\icmltitle{Answering Compositional Queries with Set-Theoretic Embeddings}

\begin{icmlauthorlist}
\icmlauthor{Shib Dasgupta}{xxx,yyy}
\icmlauthor{Andrew McCallum}{comp}
\icmlauthor{Steffen Rendle}{comp}
\icmlauthor{Li Zhang}{comp}
\end{icmlauthorlist}

\icmlaffiliation{yyy}{University of Massachusetts Amherst}
\icmlaffiliation{comp}{Google Research}
\icmlaffiliation{xxx}{Work was done as an Intern in Google.}

\icmlcorrespondingauthor{Shib Dasgupta}{ssdasgupta@cs.umass.edu}
\icmlcorrespondingauthor{Steffen Rendle}{srendle@google.com}

\icmlkeywords{Machine Learning, ICML}

\vskip 0.3in
]

\printAffiliationsAndNotice{}

\begin{abstract}
The need to compactly and robustly represent item-attribute relations arises in many important tasks, such as faceted browsing and recommendation systems.  
A popular machine learning approach for this task denotes that an item has an attribute by a high dot-product between vectors for the item and attribute---a representation that is not only dense, but also tends to correct noisy and incomplete data.
While this method works well for queries retrieving items by a single attribute (such as \emph{movies that are comedies}), we find that vector embeddings do not so accurately support {\sl compositional} queries (such as \emph{movies that are comedies and british but not romances}).
To address these set-theoretic compositions, this paper proposes to replace vectors with box embeddings, a region-based representation that can be thought of as learnable Venn diagrams.
We introduce a new benchmark dataset for compositional queries, and present experiments and analysis providing insights into the behavior of both.  We find that, while vector and box embeddings are equally suited to single attribute queries, for compositional queries box embeddings provide substantial advantages over vectors, particularly at the moderate and larger retrieval set sizes that are most useful for users' search and browsing.
\end{abstract}

\section{Introduction}
\label{Introduction}

Representing which items correspond to which attributes is a key capability in many applications. It can facilitate tasks such as user browsing, faceted search, and recommendations which are crucial in diverse platforms for helping the user discover useful content.

A popular machine learning approach for this task represents items and attributes by vector embeddings, and indicates that an item has an attribute by a high dot-product among their vectors.  An embedding representation has multiple advantages:  not only does it compactly represent the full item-attribute co-occurrence matrix, but it also generalizes beyond the training data in ways that tend to correct noisy and incomplete observations.  For example, if the set of video attribute data includes a diverse set of user-generated tags, the association of these tags will certainly be incomplete and noisy; however, users' ability to reliably browse by such diverse tags is a valuable amenity.  The embedding representation enables a movie to be beneficially labeled with tags beyond the incomplete set hand-applied by users.

A simple use-case for item-attribute relations is querying for items having a single relation, such \emph{songs in the jazz genre}.  The predominant, traditional evaluation of such systems addresses this single-attribute case.  However, there is increasing interest in accurately handling richer, multi-attribute queries representing {\sl set-theoretic compositional} queries, such as \emph{songs that are jazz but not smooth-jazz} or  \emph{movies for children that are animated but not having monsters}.  Formally, this corresponds to conjunctions of (possibly negated) attributes, e.g., \emph{jazz $\land \neg$ smooth-jazz} or \emph{children $\land$ animation $\land \neg$ monster}.
While the single attribute case is well studied, there is much less work on how to handle compositional queries in embedding spaces.
The combinations of concepts are inherently set-theoretic, which geometrically, we argue, are more naturally represented by {\sl regions} rather than vector {\sl points}.

This paper studies set-theoretic compositional attribute queries on items, introduces a new benchmark dataset, analyzes the behavior of vector embeddings on such queries, and proposes to replace vectors with box embeddings~\cite{vilnis2015word}, a region-based representation that can be thought of as learnable Venn diagrams, supporting region-based intersection and negation.

Our results and analysis shed light on what makes the compositional task so challenging:
For single concept classification, it usually suffices to focus on the ``top’’ results.
However, for compositional queries, we are forced to achieve good quality on the tail and understand the boundaries of concepts.
For instance, given a query \emph{jazz $\land \neg$ smooth jazz}, it is not sufficient to understand the top \emph{jazz} and the top \emph{smooth jazz} tracks, but the model needs to understand the boundaries of \emph{smooth jazz} to make judgments about songs.

In this paper, we first describe compositional queries represented in vector embedding spaces.
We discuss the prior work on heuristics to combine results based on a probabilistic interpretation of similarity scores, and based on vector space arithmetic~\cite{mikolov2013efficient}, and discuss the weaknesses of the vector-point-based representations.
Then we discuss various region-based set theoretic embeddings, settling on the particular advantages of box embeddings~\cite{vilnis2015word}.  Box embeddings naturally represent regions whose volume is easy to calculate, are closed under intersection, efficiently represent negation for the common case of a small number of negations, and naturally have good generalization properties.
Although the box embedding methodology has been introduced previously, this paper is the first to analyze its compositional properties empirically in depth.

A key challenge to empirical studying compositional queries is the lack of existing datasets and benchmarks for this task.
In this work, we propose a new benchmark that contains a set of compositional queries with ground truth items as well as noisy and incomplete data for training\footnote{Link to the benchmark: \burl{https://github.com/google-research-datasets/genre2movies}}.
Our raw data sources are the combination of the widely-studied MovieLens dataset~\cite{harper:2015} and Wikidata\footnote{\url{https://www.wikidata.org/}}.
We use the user-generated tagging data in MovieLens as the noisy and incomplete training data for the relationship between movies and genres. We use Wikidata to construct ground truth semantic labels for the movies. In addition, we carry out statistical analysis on the co-occurrence of labels to form meaningful compositional queries. We employ cross-verification to make sure the ground truth data is reasonably accurate and complete. 

We carry out a systematic empirical study and analysis of both vector and box embedding models on this benchmark.
Our study finds that, while that vector and box embeddings are equally suited for singleton queries, for compositional queries box embeddings provide significant accuracy gains, particularly when evaluated at the moderate and larger retrieval set sizes are most useful when users want to explore and browse query results.

\section{Related Work}
\label{Related Work}

In this section, we aim to put our work into the perspective of recent advancements of region-based representations, compositional querying of vector embeddings, and compositional generalization.

\subsection{Compositional Queries with Vector Embeddings}

It is common in machine learning to represent discrete entities such as items or attributes by vectors~\cite{bengio2013representation} and to learn them by fitting the training data. Besides semantic similarity, some have claimed that learned vectors have compositional properties through vector arithmetic, for example in empirical analysis of word2vec~\cite{mikolov2013efficient} and GLOVE~\cite{pennington2014glove}, and some theoretical analysis~\cite{levy2014neural,arora2018linear}.  However, anecdotally, many have found that the compositional behavior of vectors is far from reliable \cite{rogers-etal-2017-many}.  Our paper provides a comprehensive evaluation of vector embeddings on compositional queries, and compares the results to a region-based alternative.

\subsection{Region Based Embeddings}

Often concepts have different generality and their conjunction, disjunction and negation imply yet another concept, e.g, 'horror movies' is a broader concept than 'ghost movies'. One of the first efforts to capture this specificity/generality and the asymmetric relation between entities is made by \citet{vilnis2015word} where the authors propose to use Gaussian distributions to represent each concept, and KL Divergence as a score function. Many consequent methods are proposed based on region-based embeddings to solve this problem. \citet{hyperbolic_graph} uses hyperbolic disks, and \citet{hyperbolic_cone} use hyperbolic cones, however these are not closed under intersection nor are their intersections easily computable. \citet{order_embedding} and  \citet{poe} use an axis-aligned cone to represent an entailment relation. \citet{hard_box} extend the work of \citet{poe} by adding an upper-bound to the cone, resulting in a strictly increased representational capacity. \citet{softbox} and \citet{gumbel_box} propose improved training methods to handle the difficulties inherent in gradient-descent based training. In their work, \citet{word2box} use box embeddings to capture word semantics from large text corpora. They demonstrate that the trained box embeddings capture set theoretic semantics, e.g., 'tongue' $\cup$ 'body' is similar to 'mouth' but 'tongue' $\cup$ 'language' is similar to 'dialect'. These compositions are non-trivial for vector based embeddings. However, in their task, they could not quantify the degree to which these embeddings are capturing set semantics. In contrast, we develop an evaluation method to measure the degree of effectiveness of box embeddings to capture the semantics in compositional queries for recommendation systems.

\subsection{Compositional Generalization}
There have been many recent efforts to understand whether natural language processing systems demonstrate systematic intelligence \cite{sysgen1} which is the ability to produce concepts through different combinations of already known concepts. Many benchmark datasets have been proposed towards that extent e.g., SCAN \cite{scan}, COGS \cite{cogs}, CFQ \cite{cfg}. All these benchmarks are fundamentally based on creating train/test split such that the distribution of individual concepts in training remains the same during the test, but the distribution of compounds created by those concepts should differ as much as possible during the test phase. In this work, we rely on the semantics of the embedding space to answer compositional queries, thus unlike these benchmarks, we do not train on any compounds. Furthermore, our benchmark here focuses on incomplete and noisy database completion.
\citet{faithful_emb, sun2020guessing, query2box} are some of the recent works that focus on logical query over knowledge bases (KB).

\section{Problem Formulation}
\label{Problem Formulation}

Let $I$ be a set of $m$ \emph{items} and let $A$ be a set of $n$ \emph{attribute} descriptors of the items.
For example, $I$ could be a set of movies and $A$ a set of genres or actors.
Let $O \subseteq I \times A$ be a set of movie-attribute pairs.
Each pair $(i, a) \in O$ represents an assignment of an attribute to an item. 
For example, it could represent that a particular movie is a \emph{comedy} movie.
Alternatively, $O$ can be seen as $m \times n$ matrix with $O_{i,a} = 1 \iff (i,a) \in O$.
\vspace{-0.3cm}
\paragraph{Compositional Query}
We define a \emph{query} as a logical combination of attributes.
We denote the set of all queries as $Q$.
A \emph{singleton} query consists of a single attribute $a \in A$.
We are especially interested in the conjunctive compositional queries with the form of $q_1\land q_2 \land\cdots \land q_k$, where each $q_i$ is of the form of $a_i$ or $\neg a_i$ for some attribute $a_i$. Some examples of conjunctive queries are  \emph{intersection} queries $a_1 \land a_2$, such as \emph{`comedy and romantic'}; and \emph{difference} queries, such as \emph{`comedy but not romantic comedy'}. In our evaluation (Section~\ref{sec:evaluation}), we consider combinations of up to three attributes with at least one positive attribute.

The conjunctive queries are probably the most common and natural queries -- they also entail a clear semantics. The other main reason for focusing on the conjunctive queries is that they convey the main challenges of the compositional queries -- conjunction would refine the query so to lead to increasingly ``sparse'' results; and the negation is usually missing from the training data so as to force the model to understand the ``boundary'' of an attribute. Indeed, as we shall show later, these properties do cause difficulty to models which work well on evaluation metrics of individual attributes.
\vspace{-0.3cm}
\paragraph{Querying with Ground Truth Data}

We are interested in matching items to a query based on the movie-attribute assignments $O$.
In the case of complete annotations $O$, this is a straightforward process. 
We denote the matching function as $\I$, that maps a query to a set of items.
For a singleton query $q=a$, the matching function is
\begin{align}
    \I(a) = \{i \in I | (i,a) \in O\} . 
\end{align}
The matching function for an \emph{intersection} query is
\begin{align}
    \I(a_1 \land a_2) &= \{i \in I | (i,a_1) \in O \land (i,a_2) \in O\} \notag\\
    &= \I(a_1) \cap \I(a_2)
    \label{eqn:and_query}
\end{align}
and for \emph{difference} queries:
\begin{align}
    \I(a_1 \land \neg a_2) &= \{i \in I | (i,a_1) \in O \land (i,a_2) \not\in O\}\notag \\
    &= \I(a_1) \setminus \I(a_2)
    \label{eqn:neg_query}
\end{align}
As we can observe from equations \ref{eqn:and_query} and \ref{eqn:neg_query}, querying with compositional queries is essentially equivalent to set theoretic operations of item-sets for individual attributes.
In an ideal scenario, where the ground truth $O$ is fully observed, the matching process is well defined.
\vspace{-0.3cm}
\paragraph{Querying with Incomplete and Noisy Data}

In our work, we consider the case where the attribute assignments are only partially observed and noisy.
This is a common scenario in real-world applications.
Let us denote these noisy incomplete assignments as $O' : I \times A \rightarrow \mathbb{R}$ or equivalently in matrix notation $O' \in \mathbb{R}^{I \times A}$.
Note that here assignments are not boolean but can be real valued numbers.
Values of zero do not imply that the item does not have the attribute but could just mean that we don't know its assignment.
In general, any number $O'_{i, a}$ represents a weak and noisy indicator of the true assignment $O_{i,a}$ of $a$ to $i$.
Our work aims at developing techniques to predict the matching function $\hI : Q \rightarrow 2^I$ from the noisy data $O'$.
Instead of predicting the membership, we are usually more interested in ranking the items for a query and we will propose methods that produce membership scores $\hY : Q \times I \rightarrow \mathbb{R}$.
These scores can be used for ranking the items by increasing scores, i.e., $i_1$ is more likely to match a query $q$ than $i_2$ if $\hY(q,i_1) > \hY(q,i_2)$.

\section{Method}
\label{sec:method}

In this work, we focus on embedding based models for representing attributes and for answering compositional queries.
In these models, we embed each attribute and item in the geometric space, e.g. as a vector or a box, and use their geometry, e.g. distance or volume, to capture the semantic relationship. Embedding models are quite common and effective for answering singleton queries.
However, answering compositional queries is more challenging.
We will present approaches for compositional queries based on vector and box embeddings.
While the combination for vector embeddings will be heuristic, box embeddings fit naturally the compositional aspect because of their set theoretic nature.

\subsection{Vector Embeddings}
\label{sec: vector_baseline}

Vector based methods represent items and attributes by embedding vectors, which are learned by fitting some, typically co-occurrence based, learning objective.
In this work, we focus on the matrix factorization method for learning a vector based embedding model. As shown by \citet{pennington2014glove}, the matrix factorization method, by setting the objective values and the weights properly, can be used to produce embeddings that achieve state-of-the-art evaluation results on the word analogy task, which is related to the compositional query task.

In such a method, the incomplete and noisy matrix $O'$ can be factorized as,
\begin{equation}
    O' \approx U \, V^T 
    \label{eqn:matrix_factorization}
\end{equation}
with $U \in \mathbb{R}^{I \times d}$ and $V \in \mathbb{R}^{A \times d}$. Each row vector $u_i\in\mathbb{R}^d$ in $U$ (or $v_a\in\mathbb{R}^d$ in $V$) is the vector representation of the item $i$ (respectively attribute $a$). We would like their dot product $\langle u_i, v_a\rangle$ to be close to the observation $O'_{i,a}$, where the closeness is defined through a loss function, with more details described below.

\subsubsection{Training}

There is a large body of literature on defining the loss function for learning $U$ and $V$. The main options are on how the loss is defined for each $(i,a)$ pair and on how much weight is given to each pair. In this paper, we discuss a few broadly used methods.

First, we apply a transformation function $\Phi: \mathbb{R}\to\mathbb{R}$ to the dot product to obtain a score, i.e.
\begin{align}\label{eq:y}
    \hY(a, i) := \Phi(\langle u_i,v_a \rangle)\,.
\end{align}
Here we consider either the identity or the sigmoid function for $\Phi$. For each choice of $\Phi$, we define a loss function for measuring how ``far'' the prediction is from the observation. When $\Phi$ is the identity function, we use the hinge-loss; and when $\Phi$ is the sigmoid function, we use the cross-entropy loss.

Since our data contains only positive pairs, we apply the following common negative sampling method to create negative samples: whenever the optimization algorithm encounters a non-zero item-attribute pair $O'_{i,a}$, it also samples a few negatives by randomly changing either the item or the attribute, i.e. by setting $O'_{i',a}=0$ and $O'_{i,a'}=0$ for a randomly chosen $i'\neq i$ and $a'\neq a$. Then we learn $U,V$ by minimizing the loss through the stochastic gradient method~\cite{koren:matrixfactorization}.

Intuitively, the loss defined above encourages the embeddings of attributes and items that have non-zero values in $O'$ to come closer in terms of the dot product similarity measure, while pushing apart embeddings of those item-attribute pairs that have a value of zero in $O'$.

\subsubsection{Singleton Query}

The trained embedding model provides a natural way to answer singleton queries, i.e. for $q=a$, we can directly use the score $\hY(a,i)$ as defined by (\ref{eq:y}).
When we want to predict a boolean membership $\hI(a)$, the ranked scores can be thresholded
\begin{align}
    \mathcal{I}(a) = \{i: \hY(a,i) > \tau_a\},
    \label{eqn:single_query_vector}
\end{align}
where $\tau_a$ can be chosen dependent on the evaluation metric.

\subsubsection{Compositional Query}

\label{sec:composite_query}

Answering compositional queries with vector embedding models is more challenging.
The matrix factorization model (Equation~\ref{eqn:matrix_factorization}) does not provide a natural way to answer queries such as intersection queries $a_1 \land a_2$ or difference queries $a_1 \land \neg a_2$.
Here, we will discuss two heuristic approaches. One of them is based on score aggregation of individual attribute scores. The other is based on capturing the composition semantics in the embedding space.

\vspace{-0.3cm}
\paragraph{Score Aggregation}
One way to answer compositional queries is to combine the item scores of individual attribute scores by treating the score $\hY(a,i)$ as the probability of the item $i$ satisfying the attribute $a$. This only makes sense when $\hY(a,i)$ is defined in the range of $[0, 1]$, for example when $\Phi$ is the sigmoid function. Then naturally one can define:
\begin{align*}
\hY(\neg a, i) &= 1-\hY(a, i)\,,\\
\hY(q_1\land q_2, i) &= \hY(q_1, i) \cdot \hY(q_2,i)\,.
\end{align*}

With these formulas, we can obtain a score between any conjunctive query and any item. The formula here implicitly assumes that the attributes are independent of each other. While this serves as a reasonable heuristic, it may fail when the attributes are correlated.

\vspace{-0.3cm}
\paragraph{Embedding Aggregation}
In this approach, we rely on the embedding space semantics to generalize over the compositional semantics.
Instead of aggregating scores, we aggregate the underlying embeddings using vector arithmetic, such as summation and subtraction between vectors to represent their composition. This has been shown to be effective in practice~\cite{mikolov2013efficient,pennington2014glove} and justified in theory~\cite{levy2014neural, arora2018linear}.

Under such vector arithmetic, we use summation for the intersection queries, i.e.
\begin{equation}
    \hY(a_1 \land a_2, i) = \Phi(\langle u_i, v_{a_1} + v_{a_2} \rangle),,
\end{equation}
and subtraction for the difference query,
\begin{equation}
    \hY(a_1 \land \neg a_2, i) = \Phi(\langle u_i, v_{a_1} - v_{a_2} \rangle).
\end{equation}

\subsection{Set theoretic Embeddings}
\label{sec: box_embeddings}
We observe that the inherent nature of predicting combinations of queries is set theoretic. To illustrate further, consider the equation \ref{eqn:and_query} \& equation \ref{eqn:neg_query} which correspond to the conjunction and negation of two queries. In there, note how the conjunction can be interpreted as the intersection between the item-set retrieved for individual queries (similar for the negation as well). However, in the case of vector based embeddings, the choices to represent this set operations are not natural and do not conform naturally to the set theoretic axioms.
We briefly describe Box embeddings and then discuss how they can be used for compositional queries.

\subsubsection{Overview of Box Embeddings}
Box embeddings were first introduced by \citet{hard_box} where an elements $\mathbf a$ of some set $A$ is represented through Cartesian product of intervals,
\begin{align}
\label{eq: hard box definition}
\begin{split}
    \Box(\mathbf a) &\defeq \prod_{i=1}^d [a_i^\boxmin, a_i^\boxmax]\\
    &= [a_1^\boxmin, a_1^\boxmax]\times \cdots \times[a_d^\boxmin, a_d^\boxmax] \subseteq \mathbb R^d.
\end{split}
\end{align}
This can be thought of as a $d$- dimensional hyper rectangle in euclidean space. The volume of a box can be calculated by multiplying the side lengths of the rectangle, 
\begin{equation*}
    |\Box(\mathbf a)| = \prod_{i=1}^d \max(0, a_i^\boxmax - a_i^\boxmin),
\end{equation*}
and when two boxes intersect, their intersection is yet another box,
\begin{align*}
    \Box(\mathbf a)\cap \Box(\mathbf b) =
    \prod_{i=1}^d
    [\max(a_i^\boxmin, b_i^\boxmin), \min(a_i^\boxmax, b_i^\boxmax)].
\end{align*}
This min and max operations involved in intersection hinders gradient based training because they cause large areas of the parameter space with no gradient signal. \citet{gumbel_box} proposed GumbelBox ($\GBox$) to solve this problem. The corners of the boxes $\{a_i^\pm\}$ are replaced with Gumbel random variables, $\{A_i^\pm\}$, where the probability of any point $\mathbf{z} \in \RR^d$ being inside the box $\Box_G(\mathbf{a})$ is given by
\[P(\mathbf z \in \GBox(\mathbf a)) = \prod_{i=1}^d P(z_i > A_i^-)P(z_i < A_i^+).\]
This probability can also be thought of as soft membership function, thus Gumbel Box can also be interpreted as representation for fuzzy set \cite{word2box}. The Gumbel distribution was chosen as it was min/max stable, thus the intersection $\GBox(\mathbf a) \cap \GBox(\mathbf b)$ which was defined as a new box with corners modeled by the random variables $\{C_i^\pm\}$ where
\begin{equation*}
    C_i^- \defeq \max(A_i^-, B_i^-) \text{ and } 
    C_i^+ \defeq \min(A_i^+, B_i^+)
\end{equation*}
is actually a Gumbel box as well. This max and min over the random variable bolis down to \emph{logsumexp} over the end points, which is a smooth function. The volume function of Gumbel box is a smooth function of the parameters as well.
\begin{equation*}
    |\GBox(\mathbf a)| = \prod_{i=1}^d \text{Softplus}(a_i^\boxmax - a_i^\boxmin),
\end{equation*}
where, $\text{Softplus}(x) = \beta \log(1+\exp(\frac{x}{\beta}))$. For all further discussions, we denote Gumbelbox $\GBox$ as $\Box$.

\subsubsection{Training}

In this section, we formulate the training objective through the lens of set theoretic representation learning. Let us consider, for each attribute $a \in A$, $\Box(a)$ to be its box representation. Similarly, for each item $i \in I$, let $\Box(i)$ be its box representation. 
We conceptualize  each attribute $a$ as the set of movies that has that attribute as its descriptor. For example, the attribute "Brad Pitt" can be thought of as the set of all the movies that Brad Pitt is associated with. 

More formally, given an attribute $a$, and an item $i$, if we observe that $O'_{i,a} = 1$, then we want to the set-representation of the token $\Box(a)$ to contain the representation of  $\Box(i)$. In order to enforce such a training objective, we use the probabilistic interpretation of set containment, i.e., if set $S$ contains set $S'$, then $P(S|S') = 1$. In our formulation, if $O'_{i,a} = 1$, then
\begin{align}
    & P(a | i) = 1 \\
    & \frac{|\Box(a) \cap \Box(i)|}{|\Box(i)|} = 1
    \label{eqn:probabilistic_semantics}
\end{align}
From the above equation, we can see that we are able to calculate this conditional probability with the model parameters. Also, for a negative sample, i.e., $O'_{i,a}=0$ for a $(i, a)$, we want the opposite, i.e., $P(a | i) = 0$. We optimize for the binary cross entropy loss between $O'_{i,a}$ and $P(a | i)$:
\begin{align*}
    \mathcal{L}_{bce} &= \\
    - &\sum_{(i,a) \in I \times A} \left[ O'_{i,a} \ln P(a| i) + (1 - O_{i,a}) \ln (1 - P(a | i))\right]
    \label{eqn:bce_loss}
\end{align*}

\subsubsection{Singleton Query}

The item scoring function for a singleton query $q=a$ can be directly represented by the probabilistic semantics of set containment (Equation~\ref{eqn:probabilistic_semantics}).
We define the scoring function as
\begin{equation}
    \hY(a,i) := \frac{|\Box(a) \cap \Box(i)|}{|\Box(i)|}.
\end{equation}
Again, for predicting sets, we can just apply a threshold
\begin{equation}
    \hI(a) = \{i \in I | \hY(a,i) > \tau\}.
    \label{eqn:single_query_box}
\end{equation}

\subsubsection{Compositional Query}
As we have argued before, the vector representation for logical-and query has posed as vector average which fails to obey many set theoretic axioms.
The choice of the box embedding based representation are natural to the set theoretic task. Their intersection is an organic representation of the logical-and composition. More formally, given two attributes $a_1$ and $a_2$, the item-set that corresponds to their logical-and can be inferred as, 
\begin{align}
    \hY(a_1 \land a_2, i) = \frac{|\Box(a_1) \cap \Box(a_2)\cap \Box(i)|}{|\Box(i)|} 
    \label{eqn:and_query_box}
\end{align}
Using inclusion-exclusion, we could express the score for difference as following, 
\begin{align}
    &\hY(a_1 \land \neg a_2, i) = \notag\\ &\frac{|\Box(a_1) \cap \Box(i)| - |\Box(a_1) \cap \Box(a_2)\cap \Box(i)|}{|\Box(i)|} 
    \label{eqn:and_query_box}
\end{align}
Similarly the  exact score for more complex queries can be computed using Inclusion-Exclusion principle.

\section{Set theoretic Evaluation Benchmark}

\label{sec:evaluation}

In the previous sections we described the methodologies for addressing set theoretic queries. In order to compare these methods, we need to evaluate them against a benchmark of compositional queries. 
However, to the best of our knowledge, no benchmark has been developed to evaluate a model that intend to solve such attribute based compositional query problem.

We developed an evaluation benchmark based on the broadly used MovieLens dataset~\cite{harper:2015}. In the benchmark, we extracted movie-genre annotations $O$ for the movie domain from Wikidata and assign genre attributes to the movies.
The noisy and incomplete annotations $O'$ from which the embedding models can be trained were created from user-generated tagging data from Movielens.
We created a set of meaningful compositional queries using the annotation statistics $O$ and heuristics.
The ground truth results for the queries follows from $O$ and set operations.
The overall statistics of the query dataset can be found it Table~\ref{tbl:datastats}.
The datasets for ground truth $O$ and for the compositional queries are available at \burl{https://github.com/google-research-datasets/genre2movies}.
In the remainder of this section, we will describe the data generation process in detail.

\subsection{Training Data}
\label{sec: training_data}
The Movielens dataset provides a set of user annotated tags for each movie. These tags are used to describe a wide variety of attributes associated with the movies, for example, genre, actor, theme, short content descriptions, reviews etc. Since the viewers are only tagging the movies with few relevant tags that they think are most important, the tags for each movie are far from complete. To add to that problem, the tags are subjective which is a source of uncertainty and noise. We use this tag-vs-movie information as the incomplete noisy estimate matrix $O'$, where the set of all movies are the item set $I$ and set of tags are the attribute set $A$.
This dataset has $|I|=19,545$ items, $|A|=35,169$ attributes and $||O'||_0 = 195,878$ non-zero item-attribute pairs.
We also note that the user ratings can be useful to provide semantic grouping of movies though we are not using them in this paper.

\begin{table}[t]
    \centering
    \caption{Statistics of generated queries and assigned movies for the proposed evaluation benchmark. $\rho$ is defined in (\ref{eq:rho}).}
    \begin{tabular}{llrr}
    \toprule
    Query type&      & \#Queries & mean $\rho$\\
    \midrule
    Singleton &$a$ &  218 & 1.0\\
    Intersection &$a \land b$ & 556    & 0.142 \\
    Difference &$a \land \neg b$ & 149  & 0.785 \\
    Triple Intersection &$a \land b \land c$ &  1604 &  0.054\\
    Triple Difference &$a \land b \land \neg c$ & 302 & 0.277\\ 
    \bottomrule
    \end{tabular}
    \label{tbl:datastats}
\end{table}

\subsection{Evaluation Benchmark}
\label{sec:eval_groundtruth}

Genres are a common type of queries for movies and composing two different genres is very common place, e.g, "children's animation", "children's animation but not monster" etc. We consider "movie genres" as our queries for comparing the performance amongst different methods to prove our hypothesis empirically. 

\paragraph{Genre Extraction}

Movielens provides some genre annotations but they are very coarse, with only $19$ different genres. The user tagging data, on the other hand, is highly diverse but very noisy. So for the ground truth, we use the high quality data source of Wikidata.
Given each movie from Movielens, we query Wikidata infobox for the genre information. Wikidata has much richer genre descriptions and, when present, is quite accurate. However, it may miss many genres as it often only retain the finest category.
To solve this problem, we extract a genre-hierarchy from Wikidata and use it to populate more genre annotations. For example, if a movie has genre 'sci-fi', and we know from the genre-hierarchy that $<$'sci-fi' \textit{isA} 'fiction'$>$, then we add 'fiction' in its annotation as well.
We treat this dataset as $O$.
This dataset has $25,878$ items, $218$ genre attributes and $||O||_0 = 83,670$ non-zero pairs.
We realise that $O$ likely has still some noise and some incompleteness, but it is significantly better than the Movielens genre and tag data. We provide a detailed analysis of the incompleteness of the benchmark in Appendix \ref{app:incompleteness}. 

\paragraph{Set-theoretic Query Generation}
With the genre annotations for each movie, it is straight forward to create single attribute queries. However, creating compositional queries requires more thoughts.
When we combine two arbitrary genres, most of these combinations will not pan out to be an interesting query, for example they can come back as a (mostly) empty set, e.g. 'sports' and 'apocalyptic' does not have anything in common, or one is completely contained in the other, e.g. 'sci-fi' is contained in 'fiction'.

These examples suggest that the relative size of the query result may be used to define ``interesting queries''. Indeed, that is how we construct compositional queries in our benchmark. 
We use the co-occurrence statistics $|\I(a \land b)|$ of two attributes $a$ and $b$ to determine which of the genres to consider when composing a complex query. Intuitively, for two queries $q_1, q_2$ and their combination $q=q_1 \operatorname{op} q_2$, we consider $q$ interesting if it is meaningful, i.e when $|\I(q)|$ is relatively larger than the size of combining two random sets, and non-trivial, i.e. when $|\I(q)|$ is smaller than either $|\I(q_1)|$ or $|\I(q_2)|$.

We apply the above criteria to obtain both pairwise and triplet queries of the form of $a\land b$, $a\land \neg b$, $a\land b\land c$, $a\land b\land \neg c$. We summarize the statistics of the queries we obtain in Table~\ref{tbl:datastats} and give some examples in Table~\ref{tbl:examples} in Appendix \ref{app:dataset_example}. To illustrate the ``difficulty'' of each query, we also compute the ratio $\rho$ between the size of the result set and the minimum size of each attribute involved in the query, i.e 
\begin{equation}\label{eq:rho}
\rho(q_1\land \cdots \land q_k) = \frac{|\I(q_1 \land \cdots \land q_k)|}{\min(|\I(q_1)|, \cdots, |\I(q_k)|)}\,.
\end{equation}
The value of $\rho$ represents the chance of success if we are to randomly guess one movie from the most restrictive ``atom'' query ($a$ or $\neg a$), so it gives a sense of the sparsity of the results.

\begin{table}[]
\centering
\caption{Precision (in \%, higher the better) for the singleton queries. We select the models based on P@1 performance for each method. We observe that all the methods perform similarly for these type of queries.}
\begin{tabular}{lrrrr}
\toprule
Methods          & P@1  & P@10 & P@20 & P@50 \\ \midrule
Attribute Lookup & 34.6 & \textbf{27.8} & 24.2 & 18.7 \\
Vector           & \textbf{36.8} & 25.1 & 22.0 & 18.0   \\
Box    & \textbf{36.8} & 27.6 & \textbf{25.1} & \textbf{21.1} \\ \bottomrule
\end{tabular}
\label{tab:result_simple_queries}
\end{table}
\vspace{-0.2cm}
\section{Results}
\label{sec:results}
In this section, we use our proposed set-theoretic query benchmark to compare the methods of Section~\ref{sec:method}. 

\begin{table*}[]
\centering
\caption{Precision (in \%, higher the better) for the intersection and difference of genre query.}
\begin{tabular}{lrrrrlrrrr}
\toprule
                       & \multicolumn{4}{c}{Intersection} && \multicolumn{4}{c}{Difference} \\ \cline{2-5} \cline{7-10}
Methods                & P@1    & P@10   & P@20   & P@50  && P@1    & P@10   & P@20  & P@50  \\ \hline
Atttribute Lookup      & 24.8   & 11.5   & 7.7    & 4.1   && 44.1   & 36.0   & 31.9  & 28.4  \\
Vector (Probabilistic) & 24.1   & 15.4   & 12.6   & 9.3   && 15.3   & 11.8   & 11.3  & 10.8  \\
Vector (Algebraic)     & 19.4   & 12.6   & 11.0   & 8.5   && 33.0   & 33.1   & 32.0  & 28.4  \\
Box           & \textbf{32.9}   & \textbf{20.6}   & \textbf{16.1}  & \textbf{11.3}   && \textbf{48.4}   & \textbf{43.7}   & \textbf{43.2}  & \textbf{41.3}  \\ \bottomrule
\end{tabular}
\label{tab:result_pairs}
\end{table*}

\begin{table*}[]
\centering
\caption{Precision (in \%, higher the better) for the triple intersection and difference of genre query. }
\begin{tabular}{lrrrrlrrrr}
\toprule
             & \multicolumn{4}{c}{Triple Intersection} && \multicolumn{4}{c}{Triple Difference} \\\cline{2-5} \cline{7-10}
Methods                & P@1      & P@10     & P@20     & P@50    && P@1      & P@10    & P@20    & P@50    \\ \midrule
Attribute Lookup       & 12.2     & 3.2      & 1.7      & 0.7     && 33.1     & 17.3    & 13.3    & 8.4     \\ Vector (Probabilistic) & 15.4     & 8.4      & 6.4      & 4.4     && 11.6     & 12.2    & 11.4    & 10.8    \\
Vector (Algebraic)     & 10.9     & 7.5      & 6.2      & 4.6     && 20.1     & 18.2    & 16.5    & 13.7    \\ Box          & \textbf{20.7}     & \textbf{11.9}     & \textbf{9.0}      & \textbf{6.2}     && \textbf{36.1}     & \textbf{28.9}    & \textbf{24.8}    & \textbf{20.4}    \\ \bottomrule
\end{tabular}
\label{tab:result_triples}
\end{table*}

\subsection{Methods}
\label{sec:result_methods}
We use the movie-vs-attribute matrix, $O'$, obtained from the MovieLens dataset (more details on Section \ref{sec: training_data}) for training.
We list the baselines as well as the corresponding methods with their training details here:\\
\textbf{Attribute Lookup}: A natural way to generate a list of movies given a query is to perform a lookup in the training matrix $O'$ (see equations \ref{eqn:and_query} and \ref{eqn:neg_query} where we substitute $O$ with $O'$).
    We sort the movies based on the number of users that tagged a movie with that attribute.
    This helps to reduce tagging noise.\\
 \textbf{Vector Embedding}: We use the matrix factorization method as described in Section \ref{sec: vector_baseline} to obtain the vector representation of the attributes and movies. We denote the \textbf{Score Aggregation} technique described in \ref{sec:composite_query} as \textit{Vector (Probabilistic)} and \textbf{Embedding Aggregation} as \textit{Vector (Algebraic)} in the compositional query result (Tables~\ref{tab:result_pairs} and \ref{tab:result_triples}). \\
\textbf{Box Embedding}: We use the set-theoretic embedding method as described in \ref{sec: box_embeddings}. We train our method with containment based probabilistic semantics (equation \ref{eqn:probabilistic_semantics}). The inference for individual attributes is governed by set containment semantics, movies are ranked by the extent to which they are contained by the query attribute (equation \ref{eqn:single_query_box}). The set composition queries are handled using inclusion-exclusion of intersection scores (refer to equation \ref{eqn:and_query_box}).
\vspace{-0.2cm}
\subsection{Evaluation Protocol}
We evaluate five different types of query tasks: singleton queries ($q=a$), intersection ($q=a \land b$), difference ($q=a \land \neg b$), triple intersection ($q= a \land b \land c$) and triple difference ($q= a \land b \land \neg c$).
For every query in each query task, the methods (see Section \ref{sec:result_methods}) are queried for a ranked list of movies.
We calculate the precision@$k$ with $k \in \{1,10,20, 50\}$ of the ranked movies w.r.t the evaluation ground-truth given in our evaluation benchmark (see Section~\ref{sec:eval_groundtruth}).
We report the mean precision value over all queries in a query task. We did an extensive hyper-parameter search (see Appendix \ref{app:hyperparam} for details). The best hyper-parameter for each method is determined by using the precision@1 metric for the singleton queries ($q=a$). We use those model checkpoints to asses the performance on all other compositional queries.

\vspace{-0.3cm}
\subsection{Quantitative Results}

\subsubsection{Singleton Queries}

We report the performance of singleton queries in the Table \ref{tab:result_simple_queries}.
As can be seen, all methods perform very similarly for singleton queries.
Especially for the precision@1 metric, the results between the embedding models are very close to each other.
The quality of vector embeddings drops for larger cutoffs more than box embeddings.
This could be an artifact of hyperparameter tuning where we selected the hyperparameters for precision@1 quality (on the validation set).
Also the lookup baseline provides a good quality which indicates that sorting the movies for each genre by the frequency that the tag was assigned to a movie is effective in reducing noise.
For movies that are tagged by many users, this is equivalent to a majority vote which is a straight-forward way to resolve tagging disagreement.

\subsubsection{Compositional Queries}

Table~\ref{tab:result_pairs} shows the results for queries of pairs of attributes and Table~\ref{tab:result_triples} for attribute triples.
\vspace{-0.5cm}
\paragraph{Attribute Lookup}

The lookup baseline performs particularly poorly for intersection queries where it suffers from the incompleteness of its data $O'$ and its lack of compensating it through generalization.
This effect is especially strong for larger precision cutoffs where the lookup quality degrades quickly, e.g., the quality for intersection queries drops from $24.8\%$ at precision at 1 to $4.1\%$ for precision at 50.
Similar effects can be seen for triple intersections and triple differences.
\vspace{-0.5cm}
\paragraph{Vector Embeddings}

The embedding methods perform noticeably better on intersection queries and higher cutoffs due to their generalization capabilities.
Interestingly, the lookup baseline performs much better than vector embeddings on the difference queries that require an understanding of negation.
This holds both for pair and triple queries.
We hypothesize that this is attributed to the vector embedding's lack of understanding of the boundaries of concepts.
For example to understand when a movie is still of genre $a$ but not anymore of genre $b$.
Especially, the probabilistic treatment of vector embeddings with its independence assumptions seems to fail here.
The algebraic treatment works better for difference queries.
On the other hand, the probabilistic treatment of vector scores works slightly better for intersection queries (especially for precision at 1).
\vspace{-0.5cm}
\paragraph{Box Embeddings}

We observe that box embedding methods outperform the vector embeddings and the lookup baseline by a large margin for all compositional queries and cutoffs.
Unlike vector embeddings, box embeddings are successful for difference queries.
This indicates that box embeddings understand the boundaries of concepts better which is important for these types of queries.
Box embeddings perform well for both intersections and triples.
Unlike the attribute lookup, the quality of box embeddings also degrades slower for larger precision cutoffs, meaning that box embeddings provide better results beyond the very top of the result list than the lookup method.
The observed superior quality of box embeddings on the compositional query tasks suggests that the built-in set-theoretic semantics of box embeddings are beneficial for such query tasks.

\section{Conclusion}
\label{conclusion}

This paper has presented a study of compositional queries on item-attribute relations (especially with conjunctions and negations).
We demonstrated that this set-theoretic problem is non-trivial for embedding models, and argued that it requires not just a modeling of closeness but also a modeling of boundaries.
We have aimed to show---through discussion of representational capacity, empirical results, and analysis---the advantages of a region-based embedding, such a box embeddings, over the traditional vector embeddings.
We have also curated and released a new benchmark dataset, and proposed an evaluation protocol for movie retrieval based on compositional genres, giving an opportunity to study query intersections and differences.
In future, building on this work, we hope to extend the advantages of our approach to personalized recommendation systems that can provide personalized search.

\nocite{langley00}

\bibliography{example_paper}
\bibliographystyle{icml2023}


\appendix
\onecolumn

\section{Dataset Examples}
\label{app:dataset_example}
\begin{table*}[h]
    \centering
    \caption{Examples of generated queries for our compositional query benchmark.}
    \small
    \begin{tabular}{llll}
        \toprule
        Intersection & Difference & Triple Intersection & Triple Difference \\ 
        \midrule
film noir $\land$ sci-fi&musical $\land \neg$ romance&musical $\land$ sci-fi $\land$ thriller&drama $\land$ fiction $\land \neg$ mystery \\
crime $\land$ silent&crime $\land \neg$ drama&crime $\land$ fiction $\land$ neo-noir&adventure $\land$ treasure hunt $\land \neg$ western \\
action $\land$ fantasy&adventure $\land \neg$ sci-fi&comedy $\land$ crime $\land$ screwball comedy&fantasy $\land$ horror $\land \neg$ monster \\
adventure $\land$ documentary&experimental $\land \neg$ short&action $\land$ crime $\land$ prison&adventure $\land$ buddy $\land \neg$ comedy \\
post-apocalyptic $\land$ romance&animation $\land \neg$ family&children $\land$ comedy $\land$ fiction&drama $\land$ monster $\land \neg$ werewolf \\
    \bottomrule
    \end{tabular}
    \label{tbl:examples}
\end{table*}

\section{Hyperparameters}
\label{app:hyperparam}
We did extensive hyper-parameter search for both the vector \& box embedding method.

\subsection{Search Space}
We trained with  $\text{batch size} \in \{128, 256, 512, 1024 \}$, $\text{learning rate} \in \{10^{-5}, 10^{-4}, 10^{-3}, 10^{-2}, 10^{-1}, 1\}$, and $\text{regularization coefficient} \in \{10^{-4}, 10^{-3}, 10^{-2}\}$.
We used $100$-dimensional vector embeddings for both movies and their attributes.
Since, a box embedding has both minimum and maximum co-ordinate parameters, we use $50$-dimensional box embedding to keep the number of model parameters identical to vector embeddings for a fair comparison.
Additionally, the box embedding uses temperature parameters for intersection ($\beta$) and volume ($\tau$).
We search over  $\beta \in \{0.0001, 0.001, 0.01, 1.0\}$and $\tau \in \{0.1, 0.5, 1.0\}$.
We use a random search technique to search for the best hyper-parameter. 
\subsection{Best Hyperparameters}

\paragraph{Box embeddings} Box intersection temp=0.001	box volume temp=1, negative sample movies=50,	negative sample tags=20	learning rate=1, batch size=1024.
\paragraph{Vector based method} loss type= max margin	negative sample movies=50,	negative sample tags=20, reg coefficient = 0.001, batch size = 1024, learninig rate = 1.0.
\section{Incompleteness Analysis}
\label{app:incompleteness}
We have manually verified the accuracy of annotations on samples, and we found that a positive (movie, attribute)-pair is usually correct.
However, it is more difficult to verify the coverage, i.e. how much is missing from the data set.
We made an effort to estimate its completeness by comparing $O$ with $O'$.
Intuitively, if $O$ is almost complete, then if a genre appears in $O'$, it should also appear in $O$.
For a genre $a$, we estimate the size of the true (hidden) annotation by $\frac{||O_{\cdot,a}||_0 ||O'_{\cdot,a}||_0}{||O_{\cdot,a} \odot O'_{\cdot,a}||_0}$, i.e., the fraction between the number of items annotated in $O$ multiplied by the number in $O'$ and the number of items in the intersection of $O$ and $O'$.
We can then calculate the estimated completeness of the annotation.
When analyzing only genres that have at least 5 items in the intersection, the estimated completeness of the Wikidata genre annotation is $60\%$ -- for comparison, the completeness of tagging annotations would be 23\%.
This analysis covers 34\% of all genres and 95\% of all Wikidata annotation pairs. 
Note that the completeness analysis provides only a rough estimate of the actual completeness and made assumptions such as independence and that the data does not include noise which is a particularly strong assumption for the tagging data, $O'$.
Noise in the tagging data, $O'$, leads to an underestimate of the completeness of the Wikidata annotations, $O$.
\end{document}